\documentclass[conference]{IEEEtran}

\usepackage{amsmath}
\usepackage{graphicx}
\usepackage{tikz-cd}
\usepackage{url}
\usepackage{booktabs}
\usepackage{multirow}
\usepackage{array}
\usepackage{subcaption}
\usepackage{enumitem}
\usepackage{subcaption}

\newcommand{\starter}[1]{\vspace{1mm}\noindent {\em #1.}}

\title{The Human Factor in Data Cleaning: Exploring Preferences and Biases}

\author{%
  \IEEEauthorblockN{Hazim AbdElazim, Shadman Islam, Mostafa Milani}
  \IEEEauthorblockA{Western University, London, Ontario, Canada \\
  Email: habdelaz@uwo.ca, misla2@uwo.ca, mostafa.milani@uwo.ca}
}

\begin{document}
\maketitle

\begin{abstract}
Data cleaning is often framed as a technical preprocessing step, yet in practice it relies heavily on human judgment. We report results from a controlled survey study ($N=51$) in which participants performed error detection, data repair and imputation, and entity matching tasks on census-inspired scenarios with known semantic validity.

We find systematic evidence for several cognitive bias mechanisms in data cleaning. \emph{Framing effects} arise when surface-level formatting differences (e.g., capitalization or numeric presentation) increase false-positive error flags despite unchanged semantics. \emph{Anchoring and adjustment bias} appears when expert cues shift participant decisions beyond parity, consistent with salience and availability effects. We also observe the \emph{representativeness heuristic}: atypical but valid attribute combinations are frequently flagged as erroneous, and in entity matching tasks, surface similarity produces a substantial false-positive rate with high confidence. In data repair, participants show a robust preference for leaving values missing rather than imputing plausible values, consistent with \emph{omission bias}. 

In contrast, automation-aligned switching under strong contradiction does not exceed a conservative rare-error tolerance threshold at the population level, indicating that deference to automated recommendations is limited in this setting. 

Across scenarios, bias patterns persist among technically experienced participants and across diverse workflow practices, suggesting that bias in data cleaning reflects general cognitive tendencies rather than lack of expertise. These findings motivate human-in-the-loop cleaning systems that clearly separate representation from semantics, present expert or algorithmic recommendations non-prescriptively, and support reflective evaluation of atypical but valid cases.
\end{abstract}


\section{Introduction}

Data cleaning is a central stage of data preparation that directly affects downstream reliability and fairness. Real-world datasets contain missing values, inconsistencies, duplicates, and anomalies, requiring substantial effort in error detection, imputation, duplicate resolution, and consistency checking \cite{rahm2000data, chu2016data, dallachiesa2013nadeef, chu2013katara, moslemi2026heterogeneity, moslemi2024evaluating,beyondAccuracyImputation,zheng2019currentclean,milani2019currentclean, moslemi2024threshold}. Although many components can be supported by rules, probabilistic methods, and learning-based systems \cite{fan2012towards, rekatsinas2017holoclean, krishnan2016activeclean}, cleaning remains fundamentally human-in-the-loop: practitioners validate outputs, resolve ambiguous cases, choose repairs, and interpret model suggestions \cite{zhang2019qualitative}.

This reliance on human judgment introduces an often-overlooked source of systematic distortion: cognitive bias in data cleaning decisions. While extensive research has examined bias in downstream models \cite{gebru2021datasheets}, comparatively little attention has been paid to biases introduced earlier in the pipeline during data preparation. Cleaning decisions can irreversibly reshape data distributions, remove valid but atypical records, or introduce structured distortions that propagate into model training and evaluation, yet remain difficult to audit retrospectively.

We argue that well-established cognitive biases manifest concretely in common cleaning tasks. Rather than proposing new bias categories, we examine how known mechanisms—representativeness heuristics, framing and sensitivity to surface representation, anchoring to external cues, omission bias, and selective deference to automation—shape judgments during error detection, imputation, and record linkage. Understanding how these mechanisms translate into observable cleaning actions is essential for designing bias-aware tools and workflows.

To study this question, we conduct a controlled survey experiment covering core cleaning tasks using realistic, census-inspired scenarios with known semantic validity. The design elicits behavioral outcomes—such as false-positive flags, directional revisions after expert cues, conservative inaction under uncertainty, and conflicting entity-matching judgments under ambiguity—rather than self-reported attitudes. This enables direct measurement of bias-consistent decision patterns.

Our results provide empirical evidence that multiple bias mechanisms systematically influence cleaning behavior. Surface-level formatting differences induce false-positive error detection despite unchanged semantics. Expert cues produce anchoring effects that shift decisions beyond parity. Omission bias leads many participants to leave values missing rather than commit to plausible imputations. Representativeness heuristics drive confident rejection of atypical-but-valid records and substantial false-positive entity matches. In contrast, automation-aligned switching under strong contradiction does not exceed a conservative rare-error tolerance threshold at the population level.

These effects persist across variation in cleaning effort, professional role, seniority, and workflow structure, indicating that bias in data cleaning reflects broadly shared cognitive tendencies rather than isolated knowledge gaps. By empirically grounding these mechanisms across multiple cleaning tasks, this work motivates more transparent, cognitively informed, and bias-aware human-in-the-loop data preparation systems.
\section{Related Work}
\label{sec:relatedwork}

Data cleaning has traditionally been studied as a technical problem, with extensive work on automated methods for detecting and repairing errors using rules, probabilistic inference, and learning-based techniques \cite{rahm2000data, chu2016data, fan2012towards, rekatsinas2017holoclean}. At the same time, it has long been recognized that fully automated solutions are insufficient in many practical settings. Human input is often required to validate detected errors, resolve ambiguous cases, configure cleaning rules, select among competing repairs, and interpret suggested corrections \cite{kandel2011wrangler, krishnan2016activeclean}. As a result, modern data cleaning workflows increasingly adopt human-in-the-loop designs, where users interact closely with automated systems. Despite this shift, most existing approaches implicitly treat human judgments as reliable and unbiased, focusing primarily on optimizing system performance or reducing user effort rather than examining how human decision-making itself may systematically influence cleaning outcomes.

In parallel, a substantial body of research in machine learning, natural language processing, and human--computer interaction has documented the presence of cognitive and social biases in human annotation and decision-making tasks. Prior studies show that annotator demographics, task framing, presentation cues, and contextual information can significantly affect labeling behavior and data quality \cite{geva2019we, sap2019risk, sen2015turkers}. Behavioral research further demonstrates that mechanisms such as anchoring, framing effects, omission bias, and representativeness heuristics shape human judgments even in tasks perceived as objective or technical \cite{greenwald1998measuring, milkman2009illusions}. While these findings are well established in annotation and decision-making contexts, comparatively little work has examined how such bias mechanisms manifest during data cleaning tasks, where decisions directly modify the data distribution and may be difficult or impossible to reverse.

A smaller but growing line of work has begun to examine human behavior in data management tasks using survey-based, observational, and experimental methods. These studies analyze user decisions in contexts such as schema matching and record linkage \cite{dong2005semantically}, data validation and repair \cite{koch2021clean}, and interactive or collaborative cleaning workflows \cite{reid2021human}. At the system level, recent designs increasingly acknowledge human variability by supporting iterative repair, traceability, and feedback loops \cite{pereira2024cleenex}, or by incorporating human oversight into automated pipelines \cite{rezig2019towards}. While these approaches represent important progress, they typically focus on workflow design or system support and stop short of identifying specific cognitive bias mechanisms or empirically validating their effects across multiple core cleaning tasks.

Our work complements this literature by empirically studying how well-established bias mechanisms manifest in core data cleaning tasks (error detection, repair/imputation, and entity matching) via a controlled survey study.

\section{Human Biases in Data Cleaning}
\label{sec:biases}

Data cleaning requires repeated human judgments under uncertainty (e.g., whether to flag, repair, impute, or merge records), often based on limited context and external cues (experts, tools, or summary statistics). We therefore focus on a set of well-established cognitive biases and describe how each one can surface in common data cleaning decisions.

\subsection{Anchoring and Adjustment Bias}

Anchoring and adjustment bias refers to the tendency of individuals to rely heavily on an initial value (the anchor) when making judgments, and to revise their estimates away from that anchor only partially, resulting in insufficient adjustment even when the anchor is arbitrary or uninformative. Tversky and Kahneman originally demonstrated that numerical estimates are systematically biased toward previously considered values because adjustments fail to fully incorporate relevant evidence \cite{tversky1974judgment}. Subsequent work has confirmed the robustness of this insufficient-adjustment mechanism across domains and decision contexts \cite{kahneman2011thinking}.

In data cleaning, anchors naturally arise from existing data values, default entries, suggested corrections, or reference statistics presented to the user. Once an anchor is introduced, cleaners may begin from that value and adjust their judgment incrementally rather than independently reassessing the case from first principles. When this adjustment is insufficient, the final decision remains systematically biased toward the anchor, even when substantial deviation would be normatively warranted. This can bias numerical corrections, range adjustments, and categorical replacements toward the initially presented value.

\subsection{Automation Bias}

Automation bias describes the tendency to over-rely on automated decision aids and algorithmic recommendations, leading individuals to accept incorrect suggestions or to neglect contradictory evidence. Research in human factors and ergonomics has shown that people frequently defer to automated systems, especially under time pressure or cognitive load, even when those systems are imperfect \cite{parasuraman1997humans,parasuraman2000automation}.

In data cleaning, automation bias manifests when practitioners disproportionately trust system-generated flags, suggested repairs, or imputed values. Rather than treating automated outputs as advisory, cleaners may accept them with minimal scrutiny or fail to override them when domain knowledge suggests otherwise. This bias is particularly relevant in modern data pipelines where cleaning tools increasingly incorporate rule-based or machine-learned recommendations.

\subsection{Representativeness Heuristic}

The representativeness heuristic refers to the tendency to evaluate the plausibility of an instance based on its similarity to a mental prototype, often at the expense of base rates or statistical reasoning \cite{tversky1974judgment}. Judgments guided by representativeness are sensitive to how well a case “fits” expectations, even when atypical cases are valid and informative.

In data cleaning, representativeness influences judgments about whether a record appears plausible or erroneous. Cleaners may reject or modify values that deviate from stereotypical patterns of the domain, such as atypical income–education combinations or uncommon attribute correlations, even when such records are legitimate. As a result, rare but valid data points may be disproportionately flagged or altered because they do not conform to dominant schemas.

\subsection{Framing Effects}

Framing effects occur when different but logically equivalent descriptions of a decision problem lead to systematically different choices \cite{tversky1981framing}. In other words, how a problem is presented can change decisions even when the underlying facts remain the same. Extensive research has shown that the way options, outcomes, or goals are framed can significantly influence judgment and risk perception \cite{kahneman2000choices}.

In the context of data cleaning, framing effects arise from how tasks are described and how objectives are communicated. For example, framing a task as ``detecting errors'' versus ``completing missing information'' can shift cleaners' attention toward different types of actions and outcomes. Similarly, emphasizing potential losses from incorrect edits versus gains from improved data quality can alter intervention thresholds, even when the underlying data and constraints remain unchanged.

\subsection{Availability Heuristic}

The availability heuristic describes the tendency to judge the likelihood or importance of events based on how easily examples come to mind \cite{tversky1973availability}. Salient, recent, or visually prominent information is often overweighted relative to less noticeable but statistically relevant evidence.

In data cleaning, availability effects can be triggered by visually highlighted anomalies, recently encountered errors, or attributes that draw disproportionate attention in the interface. As a result, cleaners may overestimate the prevalence or severity of certain data issues while under-attending to less salient but equally important problems. This can skew cleaning effort toward memorable cases rather than representative ones.

\subsection{Omission Bias}

Omission bias refers to the tendency to prefer inaction over action when decisions involve uncertainty, even when action could lead to better outcomes \cite{ritov1990omission}. Closely related work on ambiguity aversion shows that individuals often avoid choices with uncertain consequences \cite{ellsberg1961risk}.

In data cleaning, omission bias manifests as a preference for leaving values unchanged or missing rather than actively imputing or modifying them. When the correctness of a potential fix is uncertain, cleaners may perceive errors of commission as more costly or blameworthy than errors of omission, leading to systematic under-intervention in the presence of missing or ambiguous data.

\section{Study Design}
\label{sec:study-design}

We conducted a survey-based study to examine how well-established human biases manifest in common data cleaning decisions. The study focuses on observable decision behavior rather than self-reported attitudes, and operationalizes bias mechanisms through structured tasks that mirror realistic sources of uncertainty, plausibility judgment, and external influence in data cleaning workflows. The design is guided by the bias definitions in Section~\ref{sec:biases} and emphasizes conservative, behaviorally grounded interpretations.

The survey is organized into two parts: a \emph{background section} (Q1--Q17) that collects participant demographics and technical background, followed by five scenario-based tasks (Q18--Q47).

Participants were presented with a sequence of short data cleaning scenarios covering three core tasks: error detection, data repair and imputation, and entity matching. Each scenario asks participants to make concrete decisions—such as flagging potentially incorrect records, imputing missing values, revising earlier judgments, or resolving record pairs—based on small tables of individual-level data. Participants also provided brief justifications and confidence ratings alongside key decisions, enabling complementary analyses of reasoning patterns and judgment stability.

\starter{Scenario 1: Error Detection with Expert Cues}
The first scenario asks participants to identify records they consider incorrect or suspicious using only their own judgment. It then introduces a second step in which an external expert flags a subset of records. The records combine attributes such as education, occupation, income, and region, and include atypical but semantically valid cases (Table~\ref{tab:scenario1_example}), such as a high-income individual with no formal post-secondary education, an undergraduate software engineer with approximately \texttt{\$112,000} annual income, or less likely job--gender combinations such as a female construction worker. This scenario primarily probes the \emph{representativeness heuristic}, because participants evaluate plausibility based on expectation-driven attribute combinations rather than internal consistency. Participants are then shown the expert's flags to support measurement of \emph{anchoring and adjustment bias}, as well as secondary \emph{availability effects}, by assessing whether they disproportionately revise or reinforce their judgments toward salient, externally highlighted records.

\begin{table}[!t]
\centering
\caption{Example records used in Scenario 1 (error detection with expert cues).}
\label{tab:scenario1_example}
\resizebox{\columnwidth}{!}{
\begin{tabular}{c c c l c l l}
\hline
ID & Age & Gender & Education & Income & Occupation & Province \\
\hline
A & 36 & Male   & Bachelor's degree & \$112k & Software Engineer & Nunavut \\
B & 52 & Male   & No certificate    & \$15k  & Janitor            & Ontario \\
C & 29 & Female & Master's degree   & \$85k  & Civil Servant      & Quebec \\
D & 43 & Female & College diploma   & \$120k & Construction Mgr.  & Yukon \\
\hline
\end{tabular}
}
\end{table}

\starter{Scenario 2: Error Detection under Formatting Variations}
The second error detection scenario presents a new table of individual records and asks participants to flag any entries they consider incorrect or suspicious. Unlike Scenario~1, this scenario introduces subtle \emph{formatting variations} in otherwise correct fields to examine whether surface presentation influences error detection. Specifically, one record reports the education value as \texttt{BACHELORS} using full capitalization, while other categorical values in the table follow standard formatting. Another record reports income as a plain numeric value (e.g., \texttt{61000}) without a currency symbol or thousands separators, while other income values use conventional monetary formatting (e.g., \texttt{\$93,000}). No external guidance or expert cues are provided. 

This scenario is designed to probe \emph{framing effects} in data cleaning decisions: whether participants interpret non-standard but valid formatting as evidence of error, despite the absence of semantic inconsistency. Elevated flagging rates for these records indicate sensitivity to surface representation and presentation cues rather than underlying data validity.

\begin{table}[!htbp]
\centering
\caption{Example records used in Scenario 2 (formatting variations in error detection).}
\label{tab:scenario2_example}

\resizebox{\columnwidth}{!}{
\begin{tabular}{c c l l l}
\hline
ID & Age & Education & Occupation & Income \\
\hline
A & 50 & BACHELORS & Research Scientist & \$120000 \\
B & 45 & Bachelor's & Chef & \$47,000 \\
C & 38 & M.Sc. & Project Manager & \$93,000 \\
D & 29 & PhD & College Instructor & 61000 \\
\hline
\end{tabular}
}

\end{table}

\starter{Scenario 3: Automation Bias in Data Imputation}
In the third scenario, we examine automation bias in the context of data repair and imputation. Participants are shown a record describing a 35-year-old female teacher with a missing education attribute and are asked to impute the most appropriate education level from four options: high school, college diploma, bachelor’s degree, or master’s degree or higher. This initial decision is made based solely on contextual attributes such as age, occupation, and income. Specifically, in the initial step (Q26), participants selected an education level for this record and reported their confidence. They were then shown an automated model recommendation indicating that \emph{High School} was the most likely value (Q30), and were asked whether they would revise their original decision. Automation bias is operationalized as agreement with, or revision toward, the automated recommendation after it is revealed, particularly when it contradicts the participant’s initial judgment.

\starter{Scenario 4: Omission Bias in Data Repair}
In the fourth scenario, we study \emph{omission bias} in data repair decisions under missingness. Participants are shown a record with a missing income field; given the context, a reasonable imputation value could be zero, reflecting the unpaid nature of volunteer work. Participants are asked whether they would prefer to \emph{fill in a value} for the missing income or \emph{leave the field blank}. Omission bias is operationalized as a systematic preference for leaving missing values unfilled, even when a plausible and semantically meaningful imputation is available.

\starter{Scenario 5: Entity Matching and Representativeness Bias}
This scenario examines representativeness bias in entity matching. Participants are presented with a pair of records containing the names \emph{Li Wei} and \emph{Wei Li}, which may appear interchangeable to individuals unfamiliar with Chinese naming conventions. Both records describe individuals of Chinese ethnicity with similar annual incomes (\texttt{\$72,000} and \texttt{\$71,000}), but they differ in occupation (\emph{HR Manager} versus \emph{People Operations Lead}). Participants are asked to decide whether the two records refer to the same individual or to two different people and to report their confidence in this decision. Representativeness bias is operationalized as incorrectly matching the two records based primarily on name similarity, despite the presence of differentiating attributes that suggest they refer to distinct individuals.

All scenarios are constructed from a subset of the Canadian Census Public Use Microdata File (PUMF), which provides realistic individual-level attributes such as age, region, occupation, education, and income.\footnote{Statistics Canada. \textit{Public Use Microdata File (PUMF), Census of Population}. \url{https://www150.statcan.gc.ca}} Rather than directly sampling raw records, we design the scenarios by selecting and composing representative attribute combinations grounded in the dataset and introducing controlled variations to elicit specific decision patterns. In particular, we construct small tables that reflect plausible real-world profiles while systematically incorporating atypical but semantically valid attribute combinations, missing values requiring imputation, formatting variations that preserve semantic correctness, and ambiguous identity cases for entity matching. The scenarios are intentionally small to support careful inspection while incorporating controlled variations such as atypical but semantically valid records, missing values, and identity-salient attributes. 

Table~\ref{tab:scenario-question-bias} summarizes how each scenario maps to the specific survey questions and the primary bias mechanisms it is designed to elicit. Some biases—most notably the representativeness heuristic—recur across multiple tasks, reflecting their pervasive role in data cleaning. 

\begin{table}[htbp]
  \centering
  \caption{Mapping between survey scenarios, covered questions, and the primary human biases each scenario is designed to elicit.}
  \vspace{-2mm}
  \label{tab:scenario-question-bias}
  \resizebox{0.48\textwidth}{!}{%
    \begin{tabular}[t]{p{2.5cm} p{1.5cm} p{3.5cm}}
      \toprule

      Scenario & Questions & Bias(es) targeted \\
      \midrule

      \multirow[t]{2}{2.5cm}{S1: Error Detection}
      & Q18--Q20 & Representativeness heuristic \\

      \cmidrule(lr){2-3}
      & Q21--Q22 & Anchoring and adjustment bias \newline
                   Availability heuristic \\
      \midrule

      S2: Formatting Variation
      & Q23--Q25 & Framing effect \\
      \midrule

      S3: Imputation (Automation Cue)
      & Q26--Q30 & Automation bias \\
      \midrule

      S4: Missing Income (Omission)
      & Q38 & Omission bias \\
      \midrule

      S5: Entity Matching
      & Q39--Q47 & Representativeness heuristic \\
      \bottomrule
    \end{tabular}%

  }
\end{table}

\section{Survey Results}
\label{sec:results}

This section reports the empirical analysis of our survey responses ($N=51$). Our analysis follows a conservative, evidence-first approach: (i) we summarize participant characteristics to contextualize the sample; (ii) we test whether each scenario exhibits a statistically detectable bias manifestation at the population level; (iii) we examine whether the overall bias signature varies across selected participant subgroups using standardized evidence profiles; and (iv) we synthesize cross-scenario takeaways. Throughout, we emphasize numerical evidence (rates, effect sizes, and exact test statistics) and treat subgroup findings as exploratory due to moderate cell sizes and the descriptive nature of cross-group comparisons.

\subsection{Participant Overview}
\label{sec:results-demographics}

We summarize participant demographics and technical background using the survey \emph{background section} (Q1--Q17). These variables provide descriptive context for interpreting the scenario-level results and for defining analysis subgroups in Section~\ref{sec:results-subgroups}. We report coarse categories and high-level summaries to keep the overview interpretable and to avoid over-interpreting small intersectional cells.

\starter{Demographics and professional context}
The cohort is primarily early-career: 54.9\% of participants were aged 21--30 ($n=28$) and 27.5\% were aged 31--40 ($n=14$). Gender identity was relatively balanced, with 56.9\% identifying as male ($n=29$) and 43.1\% as female ($n=22$). Participants reported a range of geographic locations; when grouped into broad regions, 33.3\% resided in North America ($n=17$), 21.6\% in Europe ($n=11$), and 15.7\% in Africa ($n=8$), with the remainder distributed across Asia and South America or reporting unclear locations.

The sample is concentrated in technical roles (multi-select), most commonly Data Scientist (35.3\%, $n=18$), Software Engineer/Developer (27.5\%, $n=14$), and Data Analyst/Business Analyst (23.5\%, $n=12$). Technology/Software is the most frequently reported industry (58.8\%, $n=30$), followed by Research/Academia ($n=7$) and Consulting/Professional Services ($n=6$). Participants work across a range of organization sizes, with the largest groups in medium-sized organizations (51--500 employees; 43.1\%, $n=22$) and large organizations (501--5000 employees; 29.4\%, $n=15$). Seniority levels are mixed, with mid-level individual contributors (33.3\%, $n=17$) and managers or team leads (23.5\%, $n=12$) forming the largest categories. Most participants reported prior exposure to ethics or fairness topics, either through formal training (54.9\%, $n=28$) or informal exposure (39.2\%, $n=20$).

\starter{Time allocation and data cleaning workflows}
Participants reported substantial engagement with data-related work and non-trivial time devoted to data cleaning. Using midpoints of the reported time ranges, the median proportion of data-work time spent on cleaning and preprocessing is 44\% (IQR 27\%--100\%). This indicates that, for a typical participant, nearly half of their data-related effort is devoted to cleaning tasks. Post-cleaning validation practices are common: 62.7\% of participants reported using both manual and automated checks ($n=32$), 19.6\% reported automated-only checks ($n=10$), and 15.7\% reported manual-only checks ($n=8$); only one participant reported no validation.

\starter{Solo and collaborative practices}
Current data cleaning practices vary in how work is distributed between solo and collaborative settings. In practice, 45.1\% of participants reported mostly working alone (approximately 75\% solo; $n=23$), while 23.5\% reported mostly collaborative work (approximately 25\% solo; $n=12$). When asked about ideal practice, preferences shift toward more balance: 31.4\% prefer an even split between solo and collaborative work ($n=16$), while fully solo and fully collaborative preferences remain minority positions.

Overall, the participant pool spans a range of roles, industries, regions, and workflow practices, with a strong representation of technically experienced practitioners and substantial exposure to data ethics or fairness topics. This diversity supports interpreting scenario-level patterns as reflecting broadly shared decision tendencies rather than behaviors isolated to a narrow or homogeneous subgroup.

\subsection{Evidence of Bias Across Scenarios}
\label{sec:results-bias-existence}

We evaluate scenario-level bias by pooling responses across participants and testing whether each task exhibits its pre-specified behavioral signature (Section~\ref{sec:biases}). Our objective is not to demonstrate that a majority of participants are biased. Rather, we test whether a \emph{non-trivial proportion} of participants exhibit bias-consistent behavior beyond what would be tolerable under rational evaluation.

All scenario-level analyses use exact one-sample binomial tests. Let $\tau$ denote the true population probability that a participant exhibits bias-consistent behavior in a given scenario. Let $\tau_0$ denote a theoretically justified baseline probability representing tolerable behavior under rational evaluation. The binomial test evaluates whether the observed count $k$ out of $n$ participants is plausibly generated under the assumption that $\tau \le \tau_0$. Intuitively, the null hypothesis states that the bias-consistent behavior occurs no more frequently than the baseline rate $\tau_0$. Formally, we test $H_0: \tau \le \tau_0$ vs $H_1: \tau > \tau_0$. The binomial $p$-value represents the probability, under $H_0$, of observing a result at least as extreme as the one obtained. If this $p$-value is smaller than the conventional significance level $\alpha = 0.05$, we reject $H_0$ and conclude that the observed rate exceeds the tolerable baseline.

We employ two principled baseline types:

\begin{itemize}[leftmargin=4mm,nosep]
\item {\em Majority baseline ($\tau_0 = 0.5$).} Used when testing whether one of two symmetric options is selected more than half the time (e.g., omission rather than action, revision toward rather than against expert).
\item {\em Rare-error baseline ($\tau_0 = 0.10$).} Used when bias-consistent behavior should be uncommon under careful evaluation (e.g., flagging semantically valid records, producing false matches, or deferring under contradiction).
\end{itemize}

All tests use exact binomial probabilities, appropriate for $N = 51$ and small baseline rates.

\subsubsection{Scenario 1}

Scenario~1 includes two atypical-but-semantically-valid records. Participants first flag suspicious records independently, then optionally revise after viewing expert flags.

\begin{itemize}[leftmargin=4mm,nosep]
\item \emph{Representativeness heuristic.} Representativeness bias is operationalized as flagging at least one atypical-but-valid record. Out of $N = 51$ participants, $k = 31$ did so, yielding $\hat{\tau} = 31/51 = 0.608$. Because these records are semantically valid, we test $H_0: \tau \le 0.10$. An exact one-sided binomial test yields $p < 10^{-12}$, strongly rejecting the rare-error baseline. The observed prevalence is therefore far above what would be tolerable under careful evaluation. Confidence levels were high: the median confidence among those who flagged at least one atypical-but-valid record ($n = 31$) was 5 (overall median = 4), indicating that plausibility-driven flagging was both prevalent and confidently expressed.
\item \emph{Anchoring and adjustment.} Among participants who changed at least one decision after viewing expert flags ($n = 46$), $k = 31$ revised toward the expert, yielding $\hat{\tau} = 31/46 = 0.674$. Using the majority baseline $\tau_0 = 0.5$, we test $H_0: \tau \le 0.5$. An exact one-sided binomial test yields $p = 0.013$, indicating that revisions toward the expert occur more often than parity would predict. Participants therefore show systematic anchoring toward external authority.
\item \emph{Availability (salience).} Expert-flagged records show larger increases in agreement than non-flagged records. While this pattern may partially reflect anchoring, increased salience likely contributes by drawing attention to highlighted cases. We therefore interpret availability descriptively rather than as independently identified.
\end{itemize}

\subsubsection{Scenario 2} Scenario~2 isolates framing effects induced by surface-level formatting cues while holding semantic content constant. Two of five records contain non-standard but valid formatting. Flagging either formatted record constitutes framing-consistent behavior. In total, $k = 29$ of 51 participants did so, yielding $\hat{\tau} = 29/51 = 0.569$. We test $H_0: \tau \le 0.10$. An exact one-sided binomial test yields $p < 10^{-9}$, strongly rejecting the rare-error baseline. Formatting-driven framing effects therefore occur far more frequently than would be expected under careful evaluation. Among those who flagged at least one formatted record ($n = 29$), confidence was high (median = 4, IQR $[3,5]$).

\subsubsection{Scenario 3}
\label{sec:sce3-an}

Participants selected an education level before and after exposure to an automated recommendation. Five of 51 participants revised toward the automated suggestion, yielding $\hat{\tau} = 5/51 = 0.098$. Under careful evaluation, compliance with a contradictory automated recommendation should be uncommon. Using the rare-error baseline $\tau_0 = 0.10$, we test $H_0: \tau \le 0.10$. An exact one-sided binomial test yields $p = 0.59$, indicating that the observed switching rate does not exceed the conservative 10\% tolerance threshold. The Wilson 95\% confidence interval is $[4.3\%,\,21.0\%]$, reflecting uncertainty around a minority-level effect. Switching was not driven by low confidence: among the five participants who revised, mean initial confidence was 3.8 (median = 4, range 3--5). Thus, although some participants deferred to the automated recommendation, we do not obtain statistical evidence that automation-aligned switching exceeds a conservative rare-error baseline.

\subsubsection{Scenario 4: Omission Bias in Data Repair Decisions}

Participants chose whether to leave a missing income field blank or fill in a value. Across 51 participants, $k = 34$ chose omission, yielding $\hat{\tau} = 34/51 = 0.667$. Using the majority baseline $\tau_0 = 0.5$, we test $H_0: \tau \le 0.5$. An exact one-sided binomial test yields $p = 0.012$, indicating that omission occurs more frequently than parity would predict. This reflects robust omission bias: participants disproportionately avoid committing to explicit imputations even when plausible values are available.

\subsubsection{Scenario 5: Representativeness in Entity Matching}

Despite clear occupational and salary differences, $k = 24$ of 51 participants judged the two records to refer to the same individual, giving $\hat{\tau} = 24/51 = 0.471$. Because these records were constructed to be clearly non-matching, we test $H_0: \tau \le 0.10$. An exact one-sided binomial test yields $p < 10^{-6}$, strongly rejecting the rare-error baseline. The Wilson 95\% confidence interval is $[34.0\%,\,60.5\%]$, indicating a large and statistically robust false-positive rate. Confidence levels were high in both directions (mean 4.13 for ``same'' vs.\ 4.04 for ``different,'' $p = 0.912$ for confidence comparison), suggesting that errors were not driven by uncertainty.

Taken together, these results show that representativeness, framing, anchoring, and omission biases produce statistically robust deviations from conservative rational baselines, whereas automation bias in the contradiction setting does not exceed a rare-error tolerance threshold.

\subsection{Subgroup and Demographic Analyses}
\label{sec:results-subgroups}

Beyond the population-level results, we examine whether bias magnitudes vary across participant subgroups. Rather than attributing bias to fixed demographic categories, we focus on potential moderators—time allocation, professional role, seniority, and workflow structure—that may shape how participants interpret and act on data during cleaning tasks.

\begin{figure*}[t]
  \centering

  \begin{subfigure}[t]{0.32\textwidth}
    \centering
    \includegraphics[width=\linewidth]{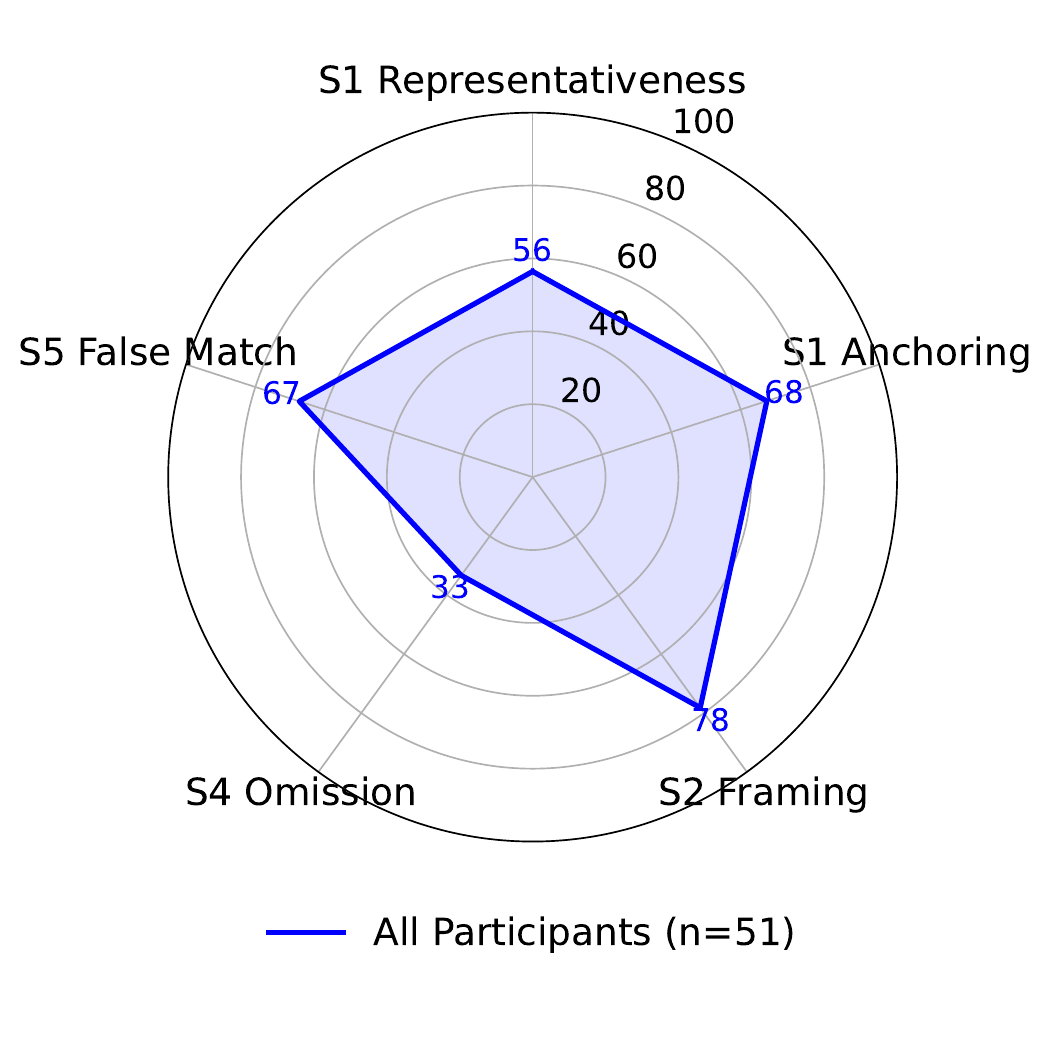}
    \caption{Overall population.}
    \label{fig:radar-overall}
  \end{subfigure}\hfill
  \begin{subfigure}[t]{0.32\textwidth}
    \centering
    \includegraphics[width=\linewidth]{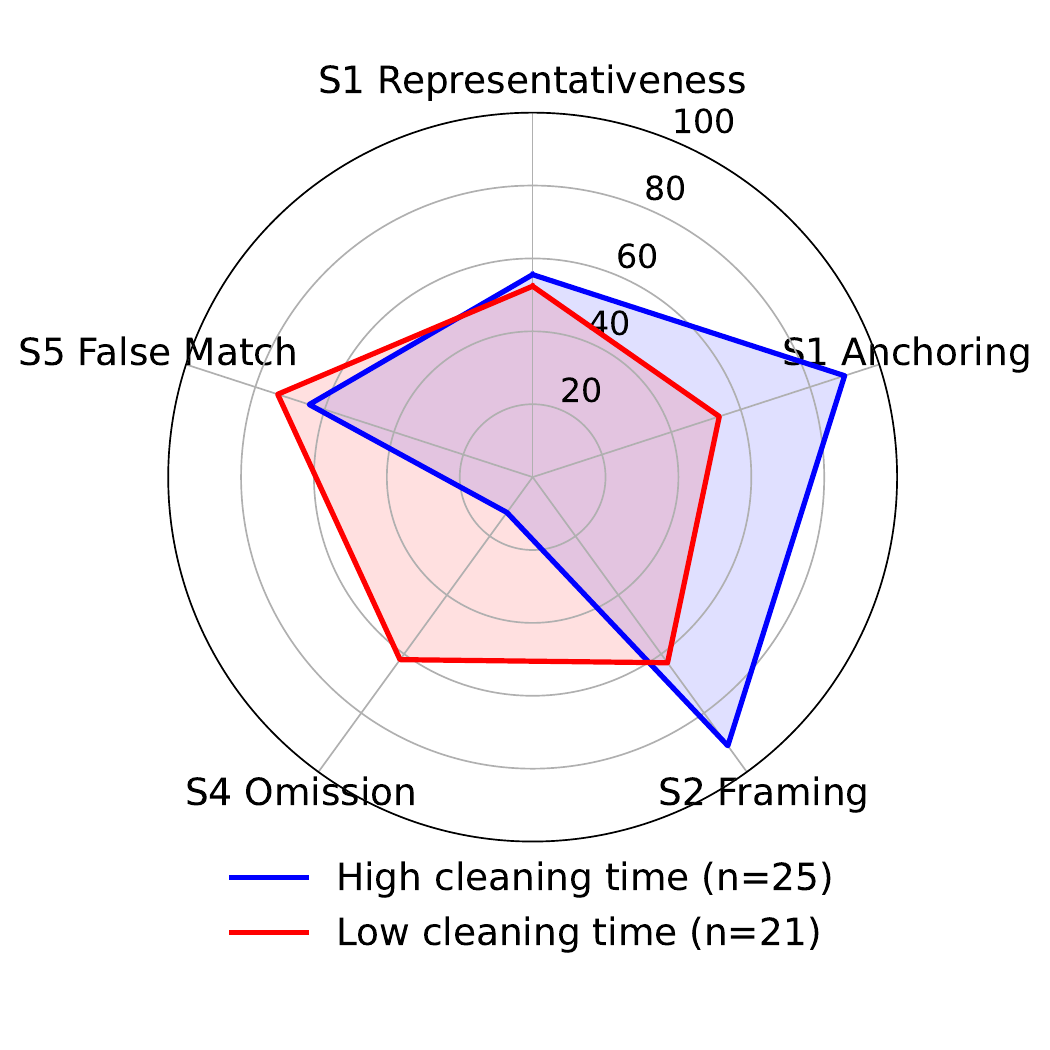}
    \caption{Higher vs.\ lower cleaning time.}
    \label{fig:radar-cleaning}
  \end{subfigure}\hfill
  \begin{subfigure}[t]{0.32\textwidth}
    \centering
    \includegraphics[width=\linewidth]{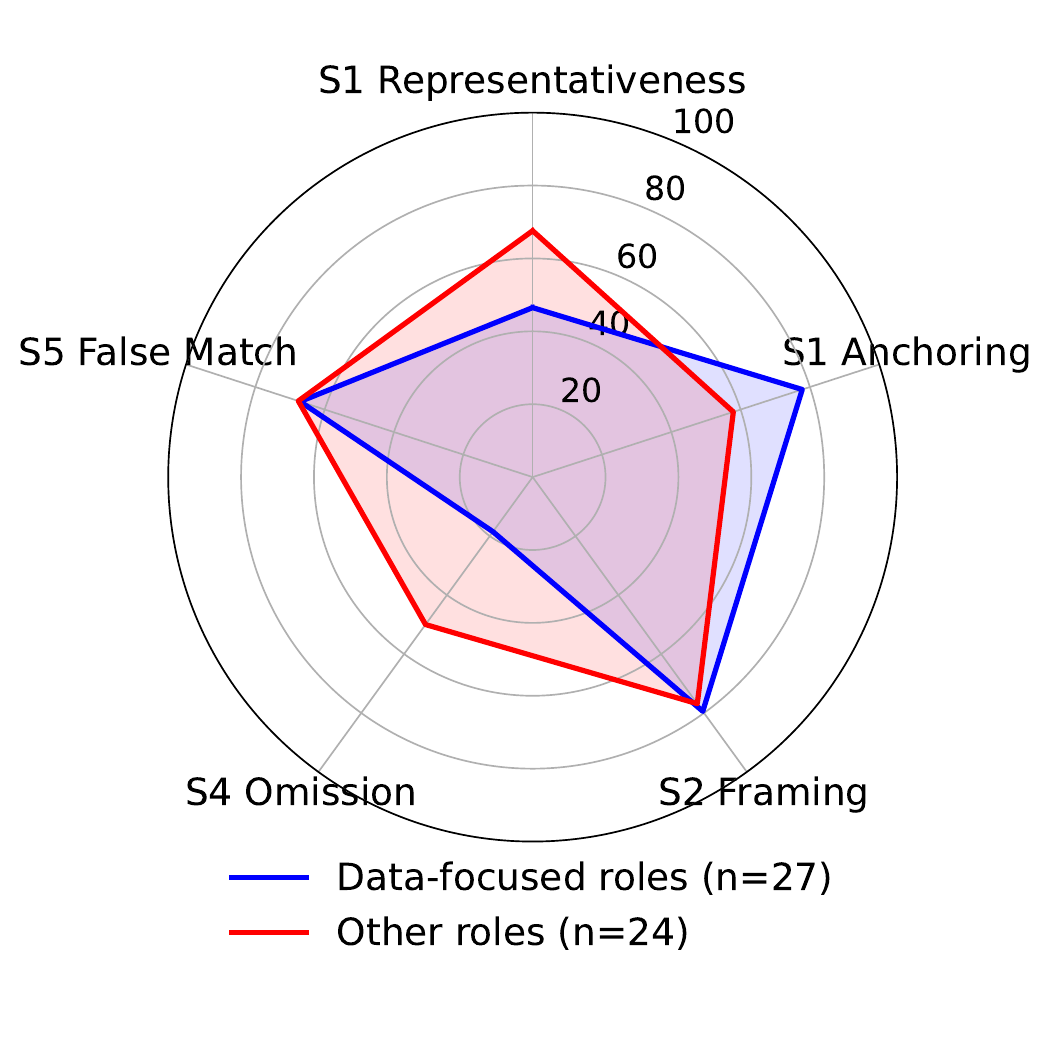}
    \caption{Data-focused vs.\ other roles.}
    \label{fig:radar-role}
  \end{subfigure}

  \vspace{0.5em}

  \begin{subfigure}[t]{0.32\textwidth}
    \centering
    \includegraphics[width=\linewidth]{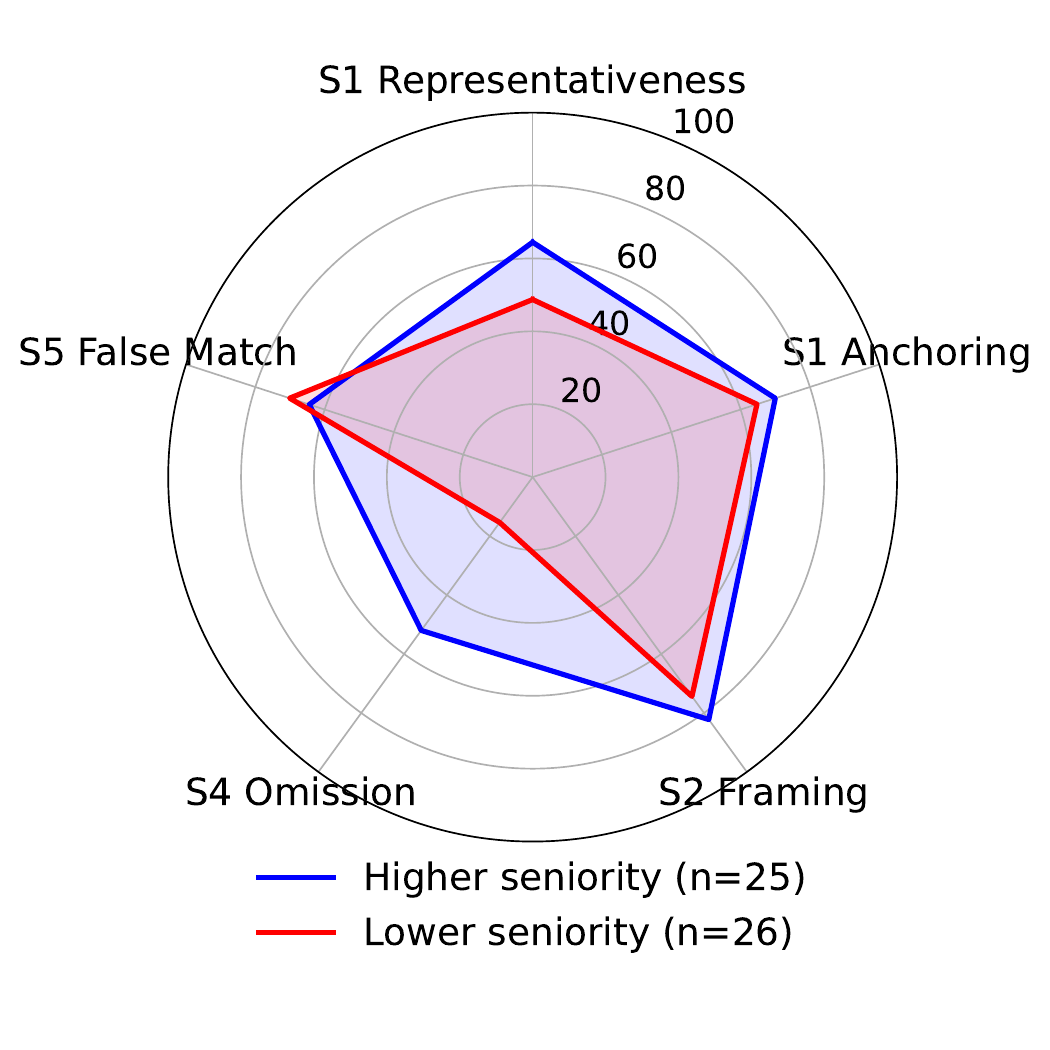}
    \caption{Higher vs.\ lower seniority.}
    \label{fig:radar-seniority}
  \end{subfigure}\hspace{1cm}
  \begin{subfigure}[t]{0.32\textwidth}
    \centering
    \includegraphics[width=\linewidth]{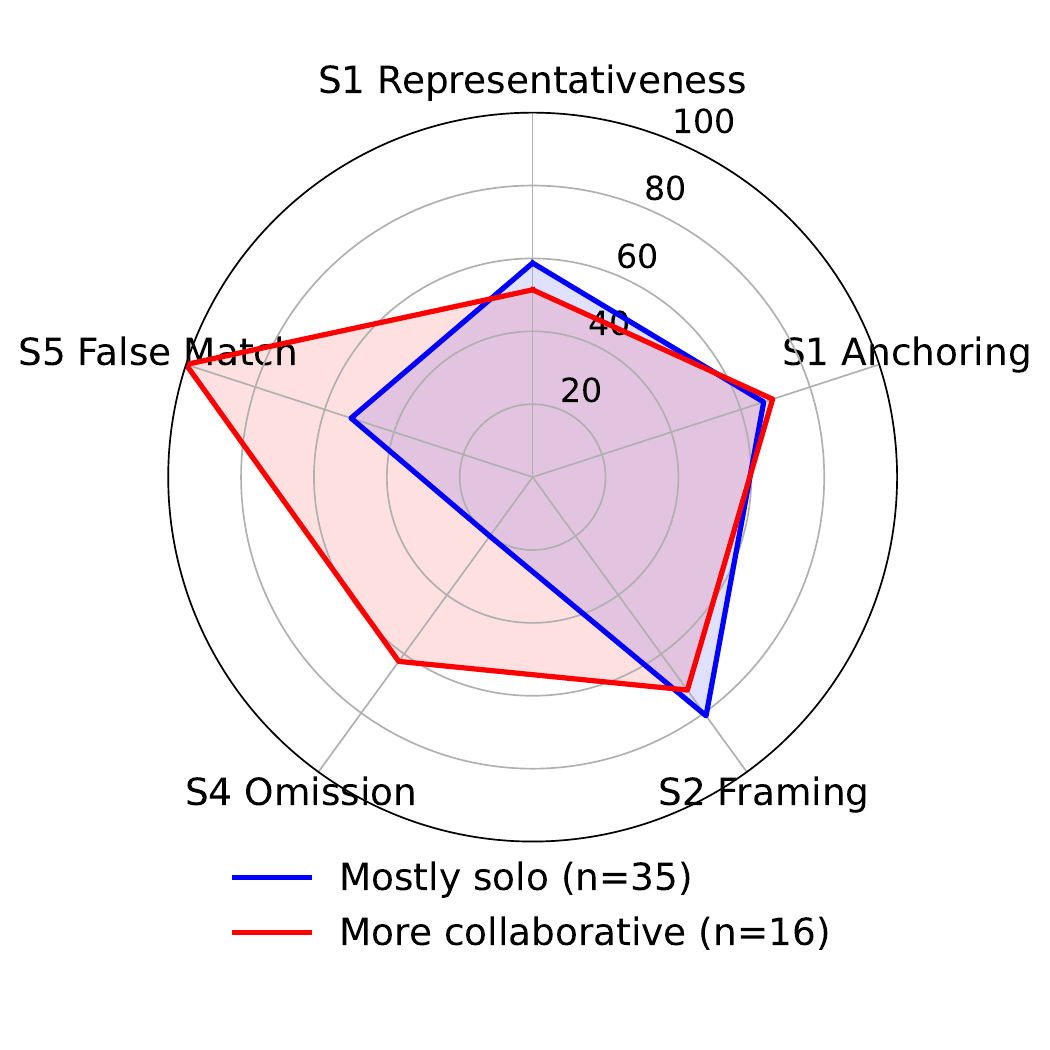}
    \caption{Mostly solo vs.\ collaborative.}
    \label{fig:radar-solo}
  \end{subfigure}

  \caption{Radar summaries of baseline exceedance scores across bias--scenario measures. Each axis reports $E=100\cdot\max\!\left(0,(\hat{\tau}-\tau_0)/(1-\tau_0)\right)$, where $\hat{\tau}$ is the observed subgroup rate and $\tau_0 \in \{0.5,0.10\}$ is the scenario-specific tolerable baseline. Larger values indicate larger behavioral deviation above the tolerable threshold (not stronger statistical significance).}
  \label{fig:radar-subgroups}
\end{figure*}

To compare subgroups across scenarios with different tolerable baselines, Figure~\ref{fig:radar-subgroups} plots a \emph{baseline exceedance score} rather than $p$-values. For each axis, let $\hat{\tau}=k/n$ denote the observed rate of bias-consistent behavior in the subgroup, and let $\tau_0\in\{0.5,0.10\}$ be the scenario-specific tolerable baseline (Section~\ref{sec:results-bias-existence}). We plot
\[
E(\hat{\tau},\tau_0)
=
100\cdot
\max\!\left(0,\frac{\hat{\tau}-\tau_0}{1-\tau_0}\right).
\]
This quantity measures how far the observed behavior exceeds the tolerable baseline, expressed as a percentage of the remaining headroom to $1$. A value of $0$ indicates that the subgroup does not exceed the tolerance; larger values indicate larger behavioral deviation above the normative threshold. We do not plot $p$-values because they quantify statistical evidence and depend strongly on subgroup sample sizes. In contrast, $E(\cdot)$ provides a directly interpretable magnitude measure. Exact binomial $p$-values are reported in Section~\ref{sec:results-bias-existence}. Scenario~3 is excluded from the radar plots because it does not exceed the rare-error baseline at the population level.

\subsubsection*{Overall profile}

Figure~\ref{fig:radar-overall} shows that framing (Scenario~2) produces the largest exceedance, followed by anchoring toward expert cues (Scenario~1) and false-positive entity matching (Scenario~5). Representativeness-driven flagging of atypical but valid records (Scenario~1) is also substantial. Omission bias (Scenario~4) appears smaller because it is measured relative to parity ($\tau_0=0.5$) rather than a rare-error tolerance ($\tau_0=0.10$). Substantively, this profile indicates that surface representation, authority cues, and similarity heuristics systematically push behavior beyond conservative rational baselines.

\subsubsection*{Time spent on data cleaning}

Figure~\ref{fig:radar-cleaning} compares participants above and below the median cleaning time. Those who spend \emph{more} time cleaning show higher exceedance in framing (Scenario~2) and anchoring (Scenario~1), whereas those who spend \emph{less} time show higher omission bias (Scenario~4). A plausible interpretation is that greater exposure to cleaning tasks heightens sensitivity to surface irregularities and external cues, amplifying framing and authority-driven effects. In contrast, participants less engaged in cleaning may default to risk-averse inaction, preferring omission over committing to explicit imputations under uncertainty.

\subsubsection*{Professional roles}

Figure~\ref{fig:radar-role} contrasts data-focused roles with other technical roles. Participants in data-focused roles exhibit lower exceedance in representativeness (Scenarios~1 and~5) and omission (Scenario~4), but higher exceedance in anchoring (Scenario~1). This pattern suggests that technical familiarity with data may reduce plausibility-based and inaction errors, yet may simultaneously increase reliance on authoritative cues. Greater comfort with structured workflows could make expert signals appear more informative, amplifying anchoring even as other biases are attenuated.

\subsubsection*{Professional seniority}

Figure~\ref{fig:radar-seniority} compares higher- and lower-seniority participants. Higher-seniority participants exhibit larger exceedance across most axes. One interpretation is that experience may strengthen confidence in heuristic judgments rather than eliminate them. Senior professionals may rely more heavily on pattern recognition and prior expectations, which can amplify representativeness and anchoring effects. This suggests that accumulated experience does not automatically neutralize bias and may, in some contexts, reinforce intuitive decision strategies.

\subsubsection*{Workflow structure: solo vs.\ collaborative}

Figure~\ref{fig:radar-solo} contrasts participants who work mostly solo with those in more collaborative workflows. Collaborative participants show higher exceedance in false-positive entity matching (Scenario~5) and omission bias (Scenario~4), whereas mostly solo participants exhibit lower exceedance across most axes. One possible interpretation is that collaborative environments may diffuse individual responsibility, making conservative inaction (omission) and similarity-based matching decisions more likely. In contrast, solo workflows may promote stronger personal accountability, leading to more cautious and self-consistent judgments. Overall, solo participants appear less biased on this normalized scale, though differences remain scenario-specific.

\subsection{Summary of Results and Implications}
\label{sec:results-synthesis}

Bias in data cleaning appears systematic and task-dependent. Representativeness heuristics drive both flagging of atypical-but-valid records (Scenario~1) and substantial false-positive entity matching (Scenario~5). Framing effects lead participants to treat surface formatting as semantic error (Scenario~2). Expert cues produce anchoring toward external authority (Scenario~1). Omission bias reflects a consistent preference for inaction over plausible imputation (Scenario~4). In contrast, automation-aligned switching under contradiction (Scenario~3) does not exceed the conservative rare-error baseline and is not treated as a robust bias signal.

Subgroup analyses indicate that these patterns are broadly shared rather than confined to specific roles or experience levels. Greater cleaning effort, data-focused roles, seniority, or collaborative workflows do not systematically eliminate bias; instead, they shift which biases are more pronounced.

These findings suggest that mitigation should focus on interface and workflow design rather than assuming expertise alone suffices. Systems should clearly separate representational irregularities from semantic errors, present recommendations as advisory rather than prescriptive, preserve reversibility for atypical cases, and encourage reflective evaluation under uncertainty. Data cleaning is therefore best understood as a cognitively mediated decision process requiring bias-aware design.
\section{Conclusion and Future Work}
\label{sec:conclusion}

This paper examined how established cognitive biases manifest in practical data cleaning decisions. Using a survey study with $N=51$ participants and controlled census-style scenarios, we showed that representativeness, framing, anchoring, and omission systematically influence error detection, imputation, and entity matching. These effects translate well-known cognitive mechanisms into concrete cleaning actions and persist across roles, experience levels, and workflow structures.

Overall, bias in data cleaning appears structural rather than incidental. Plausibility judgments override internal consistency, surface presentation shapes perceived validity, expert cues anchor revisions, and uncertainty promotes omission. Automation-aligned switching does not exceed a conservative tolerance threshold in our design, but minority susceptibility remains observable. Importantly, technical expertise or seniority does not reliably eliminate these tendencies, indicating that awareness alone is insufficient as a mitigation strategy.

\textbf{Limitations.} While our study provides controlled evidence of bias in data cleaning, it has several limitations. First, the sample size ($N=51$) is moderate, which may limit statistical power and generalizability across broader populations. Second, the scenarios are intentionally simplified and constructed using a semi-synthetic design grounded in real census data; while this enables controlled analysis, it may not fully capture the complexity, scale, and noise of real-world data cleaning workflows. Finally, participant responses may be influenced by demographic and experiential factors that are not fully controlled in this study.

\textbf{Future work.} Future work should extend this study along several directions. First, an important next step is to quantify the downstream impact of bias in data cleaning decisions, for example by measuring how bias-induced edits affect dataset quality, fairness, and model performance in subsequent learning tasks. This is non-trivial, as it requires integrating human cleaning decisions into end-to-end pipelines and isolating the effects of biased interventions from other sources of variation. One possible approach is to simulate data cleaning decisions within a standard machine learning pipeline—for example, by introducing human-like edits into a dataset, retraining models on the modified data, and comparing downstream metrics such as accuracy and fairness across different bias scenarios. Such analysis would provide a more direct link between observed human behavior and practical system outcomes.

Second, future studies should examine how these bias mechanisms appear in realistic, tool-supported data cleaning environments. Building on our controlled experimental setup, which isolates individual decisions and allows participants to revise responses after external cues, real workflows involve interactive tools where users inspect data, receive suggestions, and iteratively revise their choices. This introduces additional complexity, as decisions are shaped by evolving context, repeated interactions, and system feedback, making it harder to isolate specific bias effects. A concrete approach is to instrument existing data cleaning tools or build simple prototypes that log user actions and track how decisions evolve over time. For example, one could analyze how often users accept or override suggestions, how decisions change after repeated exposure, or whether early interactions influence later corrections. This would enable observation of bias in more realistic, multi-step workflows and better assess its impact in practice, as a methodological extension of our setup.

Finally, scaling the study to larger and more diverse populations, as well as more complex and domain-specific datasets, would improve generalizability and provide a more comprehensive view of how bias varies across different user groups. This is challenging, as it requires collecting sufficiently balanced participant samples and designing datasets that capture richer attribute dependencies and domain-specific constraints. One possible approach is to replicate the study with larger participant pools and more representative demographic distributions, while applying the same scenarios to datasets with greater scale and complexity or to domain-specific data such as healthcare or finance. This would help determine whether the observed bias patterns persist under more realistic conditions and across different populations.
\section*{AI Usage Disclosure}

AI-based tools (specifically GPT-5.3, developed by OpenAI) were used solely for language refinement tasks, including proofreading and paraphrasing to improve clarity and readability. The intellectual contributions, methodology, experiments, and conclusions are entirely the work of the authors.

AI tools were also used to assist with reference formatting. All generated content was carefully reviewed and validated by the authors.

\bibliographystyle{IEEEtran}
\bibliography{ref}

\end{document}